# Assessment of FDTD Accuracy in the Compact Hemielliptic Dielectric Lens Antenna Analysis

Artem V. Boriskin, *Member, IEEE,* Anthony Rolland, Ronan Sauleau, *Senior Member, IEEE,* and Alexander I. Nosich, *Fellow, IEEE.*

*Abstract*—The objective of the paper is to assess the accuracy of a standard FDTD code in the analysis of the near and far-field characteristics of two-dimensional (2-D) models of small-size dielectric lens antennas made of low or high-index materials and fed by the line sources. We consider extended hemielliptic lenses and use the Muller boundary integral equations (MBIE) method as a suitable reference solution. Inaccuracies of FDTD near so-called half-bowtie resonances are detected. Denser meshing reduces the error of FDTD only to a certain level determined by the type of absorbing boundary conditions used and other fine details of the code. Out of these resonances, FDTD code is demonstrated as capable of providing sufficient accuracy in the near and far-field analysis of small-size hemielliptic lenses typical for the millimeter-wave (mm-wave) applications.

## I. INTRODUCTION

Since their introduction [1, 2], integrated dielectric lens antennas (DLAs) have been recognized as an efficient and reliable solution for many mm, sub-mm, and THz band antenna applications including various indoor and outdoor communication systems [3, 4], mm-wave imaging [5], radar systems [6], collision avoidance devices [7, 8], time-domain spectroscopy [9], radio-astronomy [10], and satellite communications [11]. The integration of a dielectric lens with an active circuit has enabled one to solve two important tasks in one turn, i.e. improve the directivity of the radiating element and eliminate the losses appearing due to the excitation of surface waves on a dielectric substrate.

The most common shapes of dielectric lenses used as building blocks for integrated DLAs are hemielliptic or hemispherical. This design comes from geometrical optics (GO), which tells that all the parallel rays impinging on the elliptic lens along its major axis gather in the rear focus if the ellipse eccentricity equals the inverse of refractive index.

Later, DLAs with shaped profiles were proposed to meet specific requirements to advanced applications [11-13]. Finally, double-shell shaped lenses were introduced to improve the DLA operating bandwidth, which was restricted due to the electromagnetic performance of reduced-size conventional dielectric lenses [14, 15]. This was a step ahead in the DLA technology promising great benefits in comparison to classical hemielliptic or hemispheric designs. However, further improvement of such antennas seems to be jammed due to the lack of adequate simulation tools fast enough to be combined with optimization routines.

Among available simulation tools the most popular ones are those based on the combinations of GO and physical optics (PO) [1-4, 11-18]. They are fast and simple however provide reasonable accuracy only for electrically large lenses of low refractive index materials. As it was demonstrated in [11, 16-18], proper accounting for the multiple reflection effects improves the performance of corresponding numerical algorithms in the analysis of DLAs in the emitting mode. Recently we have found, however, that this does not prevent the same algorithms from a possible failure if applied to the analysis of the same DLAs in the receiving mode [19, 20]. This happens because the ray-tracing techniques fail to characterize accurately the focal domain size, shape, and location.

The answer to these challenges is in the use of the full-wave methods based on integral-equation (IEs) or differential-equation techniques. Among the latter ones, the most popular approach, thanks to its flexibility and simplicity of implementation, is Finite Differences in Time Domain (FDTD). It is recognized as a powerful and universal tool applicable to solving a wide variety of electromagnetic problems [21]. Unfortunately, FDTD algorithms usually have enormous requirements to computer resources especially for open-domain and/or resonant problems. This prevents such algorithms from being integrated with optimization routines. These requirements are significantly reduced for 2-D versions of FDTD [22]. Still FDTD codes have another drawback intrinsic for both 2-D and 3-D versions, which can spoil the analysis of dielectric scatterers, namely the loss of accuracy near the high-Q natural resonances [23]. This can become a bottleneck in the analysis of small-size dielectric lenses where internal resonances have been shown to play an important role [20]. Besides of [23], there seems to be no papers focused on the comprehensive verification of FDTD as to the reliable characterization of resonance phenomena in arbitrary

Manuscript received March 08, 2007; Revised Nov. 9, 2007.

This work was supported in part by the joint projects of the National Academy of Sciences of Ukraine (NASU), Centre National de la Recherche Scientifique (CNRS) and Ministère de l'Education National, de l'Eisegnement Superieur et de la Recherche, France. The first author was also supported by the National Foundation for Fundamental Research, Ukraine and by the Brittany Region via the CREATE/CONFOCAL project.

A. V. Boriskin and A. I. Nosich are with the Institute of Radiophysics and Electronics NASU, ul. Proskury 12, Kharkiv 61085, Ukraine (e-mail: a_boriskin@yahoo.com, www.ire.kharkov.ua/dep12/MOCA).

A. Rolland and R. Sauleau are with the "Groupe Antennes et Hyperfréquences", Institut d'Electronique et de Télécommunications de Rennes, Université de Rennes 1, UMR CNRS 6164, 35042 Rennes cedex, France.

dielectric scatterers. Such verification needs a trusted reference solution, and if the shape of scatterer is not circular or spherical, this can be provided by the MBIEs both in the 2-D and 3-D cases [24]. Here, convergence is guaranteed due to the favorable properties of MBIEs, and several competing algorithms for solving 2-D MBIE with controllable accuracy have been published [25-27].

The aim of this paper is to assess the accuracy of a standard FDTD algorithm in the analysis of 2-D model of compact DLA that employs a line-current fed hemielliptic lens extended with a rectangular bottom (Fig. 1).

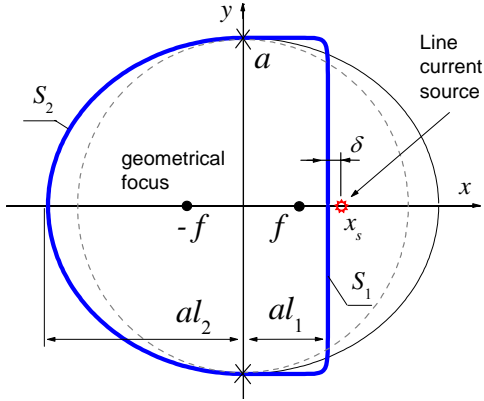

Fig. 1. Cross-section of an extended hemielliptic dielectric lens excited with a line current source.

The paper is organized as follows: a brief description of the methods used is given in Section II; then the comparative numerical data are discussed in Section III for rexolite, quartz and silicon lenses. Conclusions summarize the results obtained.

## II. OUTLINE OF THE METHODS USED

### A. Finite Differences in Time domain (FDTD)

To build our in-house FDTD algorithm, we have used a standard method with Cartesian grid [28]. The mesh cell size has been chosen to be fine enough with respect to wavelength in the lens material, $\lambda_\varepsilon$. The time has been discretized in accordance to the Courant–Friedrich–Levy stability criterion. Based on these discretizations, second-order Finite Difference approximation has been used to compute the partial derivatives in the Maxwell equations in order to establish the updated equation for each field component. The field components have been computed at successive time steps until all field intensities in the domain decay to a steady-state value. To truncate the infinite free space, absorbing boundary conditions have been implemented in the form of the split Perfectly Matched Layers (PML) [29]. As known, with a transient excitation the FDTD method provides results over a wide frequency band in one single calculation which is a benefit comparing to the frequency-domain techniques. Less known is the fact that, because of varying mesh size, the accuracy changes within the same frequency band, that is highlighted in Secion III.

In this study, to model the primary source in emitting mode we use a line current placed at the distance of one mesh cell from the lens bottom center and modulated in time with a Gaussian pulse. The central frequency is $f_0 = 40$ GHz and the duration of the pulse is chosen to provide the 20dB attenuation at $f_0 \pm 30$ GHz. The mesh size is $\Delta_x = \Delta_y = 0.1023$ mm for quartz and $\Delta_x = \Delta_y = 0.0598$ mm for silicon. The PML parameters have been set in such a way that the normal back-reflection from the boundary is kept below -50 dB for the whole frequency range.

### B. The Muller Boundary Integral Equations (MBIEs)

There are several ways to derive boundary IEs for a cylindrical dielectric scatterer. We represent the field function and its normal derivative, inside and outside the lens, as combinations of single and double layer potentials. Here, we have to assume both the contour and its normal to be continuous. Then, from the boundary conditions, we obtain a set of two coupled Fredholm second kind IEs known as the Muller boundary IEs [24].

Then we discretize MBIEs by applying the Galerkin method with entire-domain angular exponents as basis functions. The convergence rate of the algorithm is greatly improved by the application of analytical regularization in the treatment of the kernel functions [30]. This is done by adding and subtracting the canonical terms corresponding to a circular contour and analytically integrating these terms. Then the remaining parts of the matrix and right-hand-part elements are reduced to the Fourier-expansion coefficients of the smooth functions that can be economically computed with the fast Fourier transform (FFT) and double FFT (DFFT) algorithms. The resulting infinite-matrix equation has favorable features that originate from its Fredholm nature. This guarantees fast convergence of the numerical solution of matrix equation with respect to the truncation number [27].

The advantages of the developed algorithm that make it an efficient tool in the analysis of dielectric lenses are as follows: (i) controllable accuracy, i.e., a possibility to minimize the computational error for arbitrary set of lens parameters, including wavelength, lens shape and its dielectric constant, to the level determined by DFFT, by solving progressively greater matrices, (ii) low memory and time requirements, (iii) stable operation near and far from the sharp natural resonances, and (iv) absence of false "numerical resonances" intrinsic to the algorithms based on the other-type IEs [31]. More details of the algorithm properties can be found in [27].

## III. NUMERICAL RESULTS

The geometry and notations of a 2-D model of a hemielliptic expended lens antenna are given in Fig. 1. The cross-sectional contour of the lens having dielectric constant $\varepsilon$ is represented by a twice continuous curve that consists of two parts, $S_1$ and $S_2$, smoothly joined together at the points marked with crosses. Here, $S_2$ is a half of the ellipse with eccentricity chosen in accordance with the GO focusing rule, $e = \varepsilon^{-1/2}$, and $S_1$ is a half of so-called "super-ellipse" that simulates a rectangle with rounded corners [32]. In computations, we normalize lens dimensions by $k = 2\pi/\lambda_0$, where $\lambda_0$ is the wavelength in free space. This makes the obtained results valid for any frequency, thanks to scalability of the Maxwell equations. The lens is excited by a line current located close to its flat bottom. Although such a source model is idealized one, for our verification-oriented study this is not critical. More important is that such a line source can be easily incorporated into both FDTD and IEs approaches and provides an efficient excitation of resonance modes of the hemielliptic lens. As mentioned, we analyze the problem with our FDTD-2D algorithm, and the MBIE code is used as a suitable reference solution.

In Fig 2, one can see the normalized radiation patterns of the line source illuminating a quartz (left) and a silicon (right) lens with the flat bottom size $2a = 3\times\lambda_0$ that is typical for practical applications. In all computations with MBIE, we took the current line source distance to the lens bottom as $\delta = \lambda_e /10$. For comparison purposes, in the frequency scans this value was varied in the same manner for the both algorithms. As one can see, FDTD demonstrates a reasonable accuracy in this far-field analysis. However, further we will show that this is mainly due to the fact that the size of the lens is not tuned to any specific resonance.

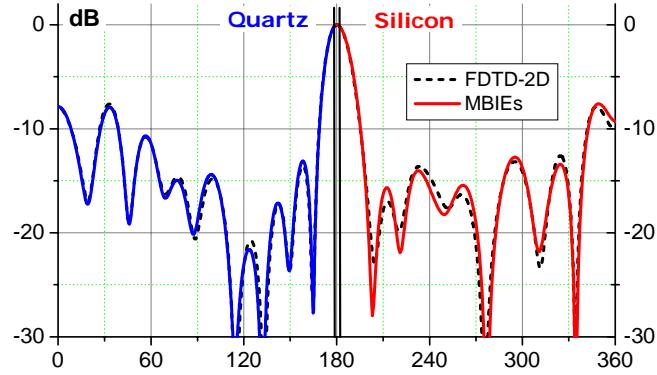

Fig. 2. Normalized radiation patterns of the electric line current illuminating quartz (left, $l_1$=0.58) and silicon (right, $l_1$=0.31) hemielliptic lenses ($ka$ = 9.42).

In the emission regime, one of the major far-field characteristic of any antenna is the directivity. For the hemi-elliptic lens antennas, on the one hand, it has been suggested that increasing the lens extension (i.e., the size of its flat-bottom side) improves its forward directivity [1, 34- 36]. In our notations of Fig 1, this is the directivity in the negative-$x$ direction. On the other hand, in our recent analysis of the same lens in the reception regime [20] we have demonstrated that this may also lead to the excitation of intensive resonances, which can dramatically affect the lens focusing ability.

The nature of these resonances has been recently revealed in [33] in the analysis of natural modes of stadium-shape and elliptic dielectric resonators. They appear as algebraic sums or differences of certain modes of the circle perturbed by the appearance of eccentricity. Characteristic field patterns of these resonances suggested their name as "bowtie" modes. Therefore, in our hemielliptic DLAs the observed resonances with triangular field patterns are on the "half-bowtie" (HBT) modes. Their $Q$-factors depend on the refractive index and electrical size and reach hundreds for a-few-wavelength lens made of dense material such as silicon.

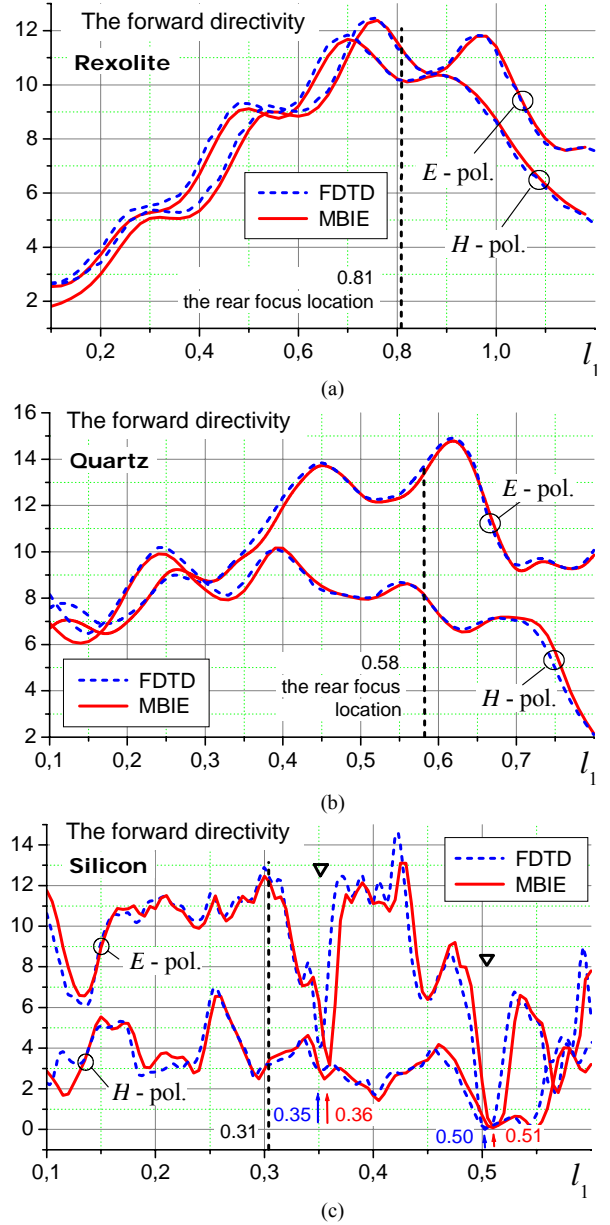

Fig. 3. The forward directivity of electrical and magnetic line currents illuminating hemielliptic lenses ($ka$ = 9.42) made of (a) rexolite ($\varepsilon$ = 2.53), (b) quartz ($\varepsilon$ = 4.0), and (c) silicon ($\varepsilon$ = 11.7) vs. the lens extension parameter. The vertical dashed lines indicate the location of the rear focus of the ellipse.

To study this effect in the emission regime and to test the accuracy of the FDTD algorithm as to description of the internal resonances, we have plotted the forward directivity of the line source illuminating rexolite, quartz and silicon lenses vs. the normalized lens extension parameter, $l_1$ (Fig. 3); both *E*- and *H*-polarization states are considered. As one can see in Fig. 3a, the rexolite lens demonstrates almost quasi-optical behavior, i.e. no significant difference between polarizations is observed and directivity grows almost monotonically until $l_1$ approaches to the rear focus value suggested by GO (indicated by the vertical dashed line). Whereas for quartz and silicon lenses the curves for different polarizations behave in different manner, and periodic oscillations (for quartz) or well defined resonance drops in the directivity (for silicon) are observed. These effects evidence a significant role of internal resonances in the lens behavior. The most important observation is that FDTD algorithm looses its accuracy near the resonances. Similarly to the previously reported FDTD analysis of the back-scattering from a circular dielectric scatterer [23], there appears a shift of resonances in frequency and also a distortion of the resonance curve.

The near-field maps for the silicon lens with extensions corresponding to the GO focus at the flat bottom and to the resonances marked with triangles in Fig. 3c are presented in Fig. 4. In Figs. 4a and 4b, one can see irregular field-spot patterns, whereas in Figs. 4c and 4f characteristic regular resonance patterns can be identified. They differ between themselves in the number of in-resonance near-field variations along a certain triangular contour marked by the dashed white line in Fig. 4c. Comparing the performance of MBIE and FDTD algorithms, one can see that, apart of the shift in frequency, FDTD is capable of reproducing the near fields of medium-*Q* resonances such as HBT ones excited within the silicon lens.

The in-resonance far-field normalized radiation patterns are presented in Fig. 5. As one can see, the resonances result in the degradation of the main lobe and even it's splitting to two side-lobes. This explains the resonance drops in directivity in Fig. 3 and agrees well with the near-field maps in Fig. 4 computed for the same resonances.

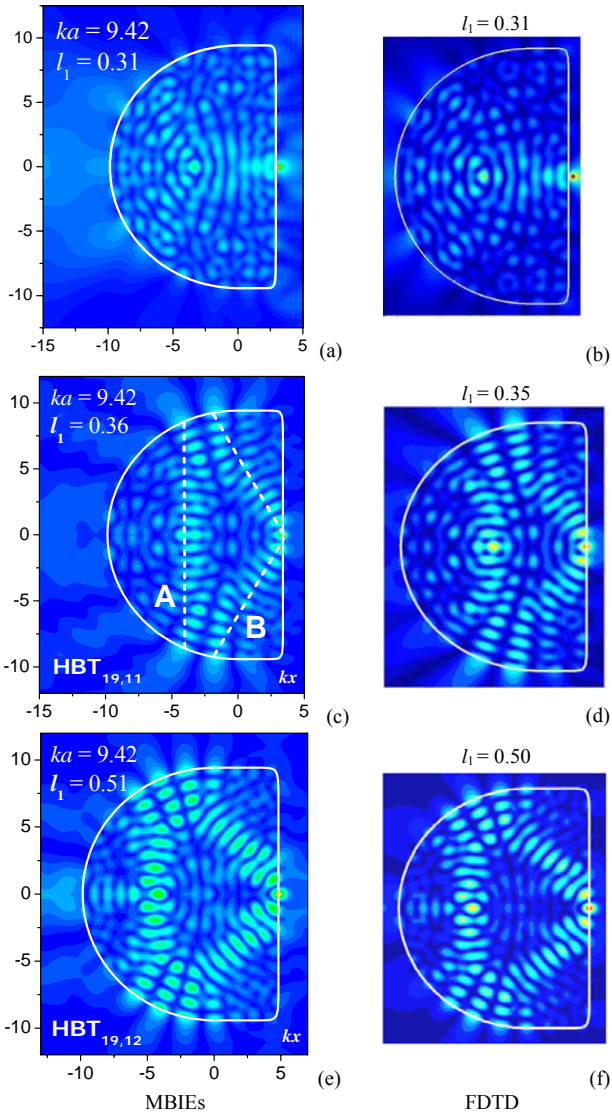

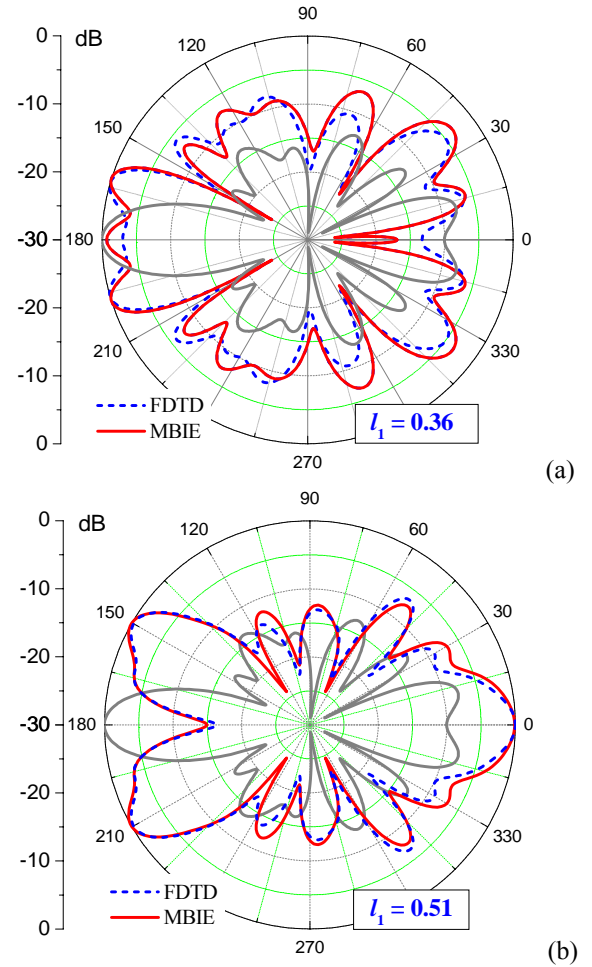

Fig. 4. Normalized near-field intensity maps of extended hemielliptic silicon lenses ($ka = 9.42$) excited by electric line currents. The lens extension values correspond to the resonances indicated by triangles in Fig. 3c. Solid white line is for the lens cross-section; dashed line in (c) highlights the characteristic triangular shape of the HBT resonance.

Fig. 5. The normalized radiation patterns of the electric line current illuminating silicon hemielliptic lenses with different resonant extensions marked by triangles in Fig. 4c. For comparison, grey line is for the radiation pattern of the line source illuminating the lens cut through the rear GO focus ($l_1 = 0.31$) computed by MBIEs (the one presented in Fig. 2 right).

Similar resonance drops in directivity appear when scanning the performance of lens in frequency – see Fig. 6. Here we computed the forward directivity of the fixed-geometry extended hemielliptic lens cut through the rear GO focus and fed by an electric line source. Comparison of the FDTD and MBIE results shows considerable discrepancies that become more pronounced in the higher frequency range. This is apparently because the mesh is getting coarser in this range as illustrated by the dotted line and the right vertical scale. The mesh size for each frequency point on the FDTD curve can be estimated as $\lambda_e/M$.

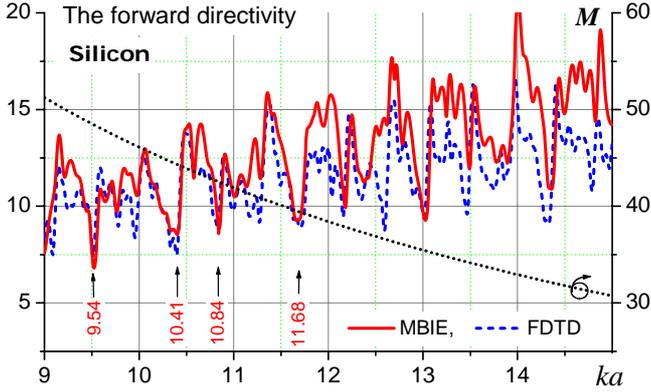

Fig. 6. The forward directivity of the electric line current illuminating a hemielliptic silicon lens (left axis) and the FDTD mesh size parameter (right axis) vs. the normalized frequency. Arrows indicate well-defined resonances whose near fields are presented in Fig. 7. The lens extension is chosen in accordance with the GO focusing rule ($l_1$ = 0.31).

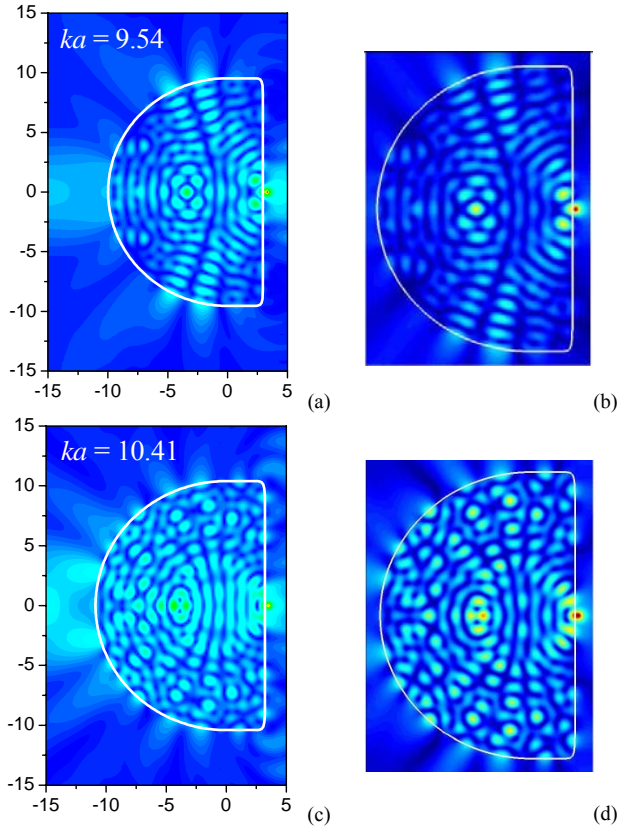

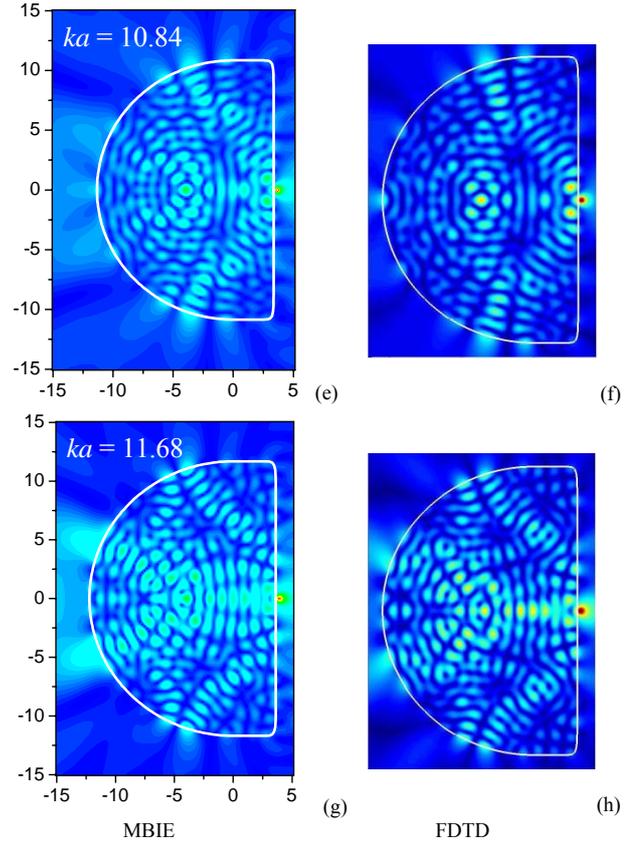

Fig. 7. Normalized near-field intensity maps of the silicon lens ($l_1$ = 0.31) whose parameters correspond to the resonances marked with arrows in Fig. 6.

Near-field maps given in Fig. 7 and the periodicity of the sharp drops in directivity confirm that these resonances are the same HBT ones as observed in Fig. 4. Moreover, one can see that the resonance presented in Fig. 7a is apparently the same as the one shown in Fig. 4c.

## IV. CONCLUSIONS

We have developed FDTD and MBIE numerical algorithms and applied them to the comparative analysis of the near and far fields of 2-D models of small-size extended hemielliptic lenses with dielectric constants corresponding to rexolite, quartz and silicon. The MBIE algorithm data have been used as a reference solution.

It has been shown that the line current of either polarization located at the axis of symmetry of the lens close to its flat bottom efficiently excites so-called HBT resonances which cause significant changes in far-field radiation pattern and, as a result, a drop in the directivity. Our comparative analysis has demonstrated satisfactorily high accuracy of the FDTD code when applied out of HBT resonances, whereas near such resonances the error in FDTD simulations can be unacceptably high. Finer meshing reduces these errors only to a certain level however cannot eliminate it completely. In particular, the resonances remain always shifted in frequency. The limiting level of accuracy is determined apparently by the type of absorbing boundary conditions used, shape and size of computational window, and other fine details of the FDTD code. Therefore these widely spread tools of numerical simulation should be used with a caution when applied in the analysis of resonance dielectric objects such as small and medium size lenses made of high-index materials.

ACKNOWLEDGMENT

The authors are grateful to Dr. S.V. Boriskina for many valuable discussions. The authors would like to thank CNRS/IDRIS, Orsay, France for the access to the high-performance computational platform.

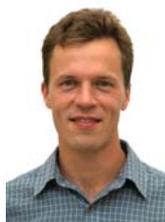

**Artem V. Boriskin** (S'99-M'04) was born in Kharkiv, Ukraine, in 1977. He received the M.S. degree in radio physics from the Kharkov National University in 1999 and Ph.D. degree in radio physics from the Institute of Radiophysics and Electronics of the National Academy of Sciences of Ukraine (IRE NASU) in 2004.

Since 2003 he has been on research staff of the Department of Computational Electromagnetics, IRE NASU. Since 2004 he has held CNRS short-time visiting postdoctoral positions at the Institute of Electronics and Telecommunications of Rennes (IETR), University of Rennes 1, France. His research interests are in development of numerical algorithms for analysis and optimization of arbitrary-shaped dielectric scatterers with application to dielectric antennas and lenses.

Dr. Boriskin received several awards including IEEE MTT-S Graduate Student Fellowship Award in 2000, student fellowship of NATO and Turkish Council of Science and Technology in 2001, 1st Prize of the European Microwave Association for the best young scientist presentation at the MSMW-04 Symposium in 2004, and the Young Scientist Research Grant of the President of Ukraine in 2006.

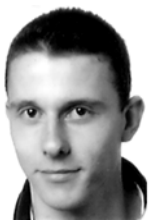

**Anthony Rolland** was born in Rennes, France in 1982. He received the Electronic Engineering degree and the Master degree in electronics from the Institut National des Sciences Appliquées, Rennes (INSA), France, in 2005.

Since 2005, he has been working toward the Ph.D degree in signal processing and telecommunications at the IETR, University of Rennes 1, Rennes, France. His main fields of interest include the analysis and optimization of lens antennas for mm-wave applications, using the FDTD technique.

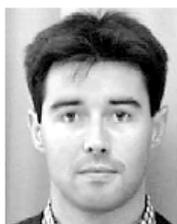

**Ronan Sauleau** (M'04-SM'06) received the Electronic Engineering and Radiocommuni-cations degree and the French DEA degree in electronics from INSA, Rennes, France, in 1995, the aggregation from Ecole Normale Supérieure de Cachan, France, in 1996, and the Doctoral degree in signal processing and telecommunications from IETR, University of Rennes 1, Rennes, in 1999.

Since 2000, he has been an Assistant Professor at the University of Rennes 1. He was elected as an Associate Professor in 2005. His main fields of interest are millimeter wave printed antennas, focusing devices, and periodic structures (electromagnetic bandgap materials and metamaterials).

Dr. Sauleau received the first Young Researcher Prize in Brittany, France in 2001 for his work on gain-enhanced Fabry-Perot antennas and was the recipient of the 2004 ISAP Symposium Young Scientist Travel Grant. In Sept. 2007, he was elected a Junior Member of the Institute Universitare de France.

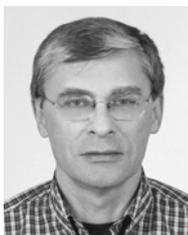

**Alexander I. Nosich** (M'94–SM'95–F'04) was born in Kharkiv, Ukraine, in 1953. He received the M.S., Ph.D., and D.Sc. degrees in radio physics from the Kharkov National University in 1975, 1979, and 1990, respectively. Since 1979, he has been with IRE NASU, Kharkov, where he is currently a Professor and Leading Scientist.

Since 1992, he has held a number of guest Fellowships and Professorships in the EU, Japan, Singapore, and Turkey. His research interests include the method of analytical regularization, propagation and scattering of waves in open waveguides, simulation of semiconductor lasers and antennas, and the history of microwaves.

Dr. Nosich was one of the initiators and a technical committee chairman of the international conference series on Mathematical Methods in Electromagnetic Theory (MMET) held in the USSR and Ukraine in 1990-2006. In 1995, he organized an IEEE AP-S Chapter in East Ukraine, the first one in the former USSR. From 2001 to 2003, he represented the Ukraine, Poland, and the Baltic States in the European Microwave Association.